\documentclass[aps,prb, preprint, citeautoscript, showpacs,floatfix,superscriptaddress]{revtex4-1}
\usepackage{bm}
\usepackage{amssymb}
\usepackage{graphicx}
\usepackage{amsmath}
\usepackage{textcomp}
\usepackage{multirow}
\usepackage{ulem}

\begin{document}
\title{Optical Coherent Injection of Carrier and Current in Twisted Bilayer graphene}
\author{Ze Zheng}
\affiliation{GPL Photonics Lab, State Key Laboratory of Applied Optics, Changchun Institute of Optics, Fine Mechanics and Physics, Chinese Academy of Sciences, Changchun, 130033, China.} 
\affiliation{University of Chinese Academy of Science, Beijing 100039, China.} 
\author{Ying Song}
\affiliation{GPL Photonics Lab, State Key Laboratory of Applied Optics, Changchun Institute of Optics, Fine Mechanics and Physics, Chinese Academy of Sciences, Changchun, 130033, China.} 
\affiliation{University of Chinese Academy of Science, Beijing 100039, China.} 
\author{Yu Wei Shan}
\affiliation{GPL Photonics Lab, State Key Laboratory of Applied Optics, Changchun Institute of Optics, Fine Mechanics and Physics, Chinese Academy of Sciences, Changchun, 130033, China.} 
\affiliation{University of Chinese Academy of Science, Beijing 100039, China.} 
\author{Wei Xin}
\affiliation{Key Laboratory of UV-Emitting Materials and Technology, Ministry of Education, Northeast Normal University, Changchun 130024, China.}
\author{Jin Luo Cheng}
\email{jlcheng@ciomp.ac.cn}
\affiliation{GPL Photonics Lab, State Key Laboratory of Applied Optics, Changchun Institute of Optics, Fine Mechanics and Physics, Chinese Academy of Sciences, Changchun, 130033, China.} 
\affiliation{University of Chinese Academy of Science, Beijing 100039, China.} 
\begin{abstract}
 We theoretically investigate optical injection processes, including one-
 and two-photon carrier injection and two-color coherent current
 injection, in twisted bilayer graphene with moderate angles. The
 electronic states are described by a continuum model, and the spectra
 of injection coefficients are numerically calculated for different
 chemical potentials and twist angles, where the transitions 
 between different bands are understood by the electron energy
 resolved injection coefficients. The comparison with the injection in
 monolayer graphene shows the significance of the interlayer coupling
 in the injection processes. For undoped  twisted bilayer graphene,
 all spectra of injection coefficients can be divided into three
 energy regimes, which vary with the twist angle. {For very low photon
   energies in the linear dispersion regime, the injection is similar
   to graphene with a renormalized Fermi 
 velocity determined by the twist angle; for very high photon energies
 where the interlayer coupling is negligible, the injection is the
 same as that of graphene; and in the 
 middle regime around the transition energy of the Van Hove
 singularity}, the injection shows fruitful fine
 structures. Furthermore, the two-photon carrier injection diverges
 for the photon energy in the middle regime due to the existence of
 double resonant transitions.
\end{abstract}
\maketitle
\section{Introduction}

In recent years, twisted bilayer graphene(TBG) has attracted great
attention in condensed matter physics as a novel platform for
studying strong correlated
phenomena
\cite{cao2018unconventional,cao2018correlated,mogera2020new,nimbalkar2020opportunities,andrei2020graphene},
topological properties, \cite{sharpe2019emergent,sharpe2021evidence}
chiralities, \cite{kim2016chiral,morell2017twisting} and nonlinear Hall
effects\cite{zhang2020giant}. The underlying physics arises from flat
bands at certain ``magic angles'', implying strong carrier-carrier
interactions. TBG is formed by the relative rotation of two {monolayer graphene} at a twist angle
$\theta$ \cite{dos2007graphene,sboychakov2015electronic}. After the rotation, the Dirac cones of these two layers
intersect and form two saddle 
points, which further lead to the Van Hove singularities (VHS) in its
density of states (DOS) or the joint density of states (JDOS)
\cite{li2010observation,yan2012angle}. At the 
``magic angles'', these two VHS merge to give flat bands
\cite{bistritzer2011moire,andrei2021marvels}. The twist angle provides an additional degree of freedom to control the band structure,
as well as the energy of VHS. For large
twist angles, the band structure lower than VHS is mostly linear, similar to that of graphene but with a smaller Fermi velocity
\cite{tabert2013optical,yin2015landau} which is determined by the twist
angle. 

The optical
properties of TBG can be effectively tuned by the twist angle, and the 
optical transitions occuring around VHS are greatly enhanced, giving featured optical
conductivity\cite{moon2013optical,tabert2013optical,yu2019gate} and
enhanced photoluminescence \cite{patel2019stacking} for optoelectronic
applications. The nonlinear optical properties of TBG have been extensively
studied for second harmonic
generation\cite{yang2020tunable,brun2015intense}, 
third harmonic generation \cite{ha2021enhanced}, third-order conductivity
\cite{zuber2021nonlinear}, high-harmonic
generation\cite{ikeda2020high,du2021high}, and 
nonlinear magneto-optic properties \cite{liu2020anomalous}. Compared
to that of graphene, the huge nonlinear conductivity due to the
resonance at VHS occurs at much lower photon energies, which might
enable possible applications on nonlinear photonic devices at long
wavelength. It is interesing to consider if other nonlinear optical
phenomena can also benifit from such VHS enhanced transitions. 

In this work, we focus on the two color optical coherent injection of
carriers and currents in TBG with 
twist angles limited in the range of  $3^\circ$ to $10^\circ$, which
are suitable for the continuum model we adopted and are easy to be
numerically evaluated. Two color optical injection 
is a third order nonlinear optical process, which utilizes
the quantum interference between the optical excitation paths of the one-photon absorption by
a weak light at frequency  $2\omega$ and degenerate two-photon absorption by 
a strong light at frequency $\omega$. It can provide a full optical way to inject
carriers and currents for studying their dynamics. Extensively
investigations have been done for bulk semiconductors\cite{atanasov1996coherent,hache1997observation,bhat2005excitonic,cheng2011two},
topological materials\cite{muniz2014coherent},
transition metal dichalcogenides\cite{cui2015coherent}, graphene\cite{sun2010coherent,rioux2011current,rioux2014interference}, and bilayer graphene\cite{rioux2011current}. The
difference between the results in graphene and bilayer graphene shows that the interlayer
coupling have strong effects on the optical injection. In TBG, because the
interlayer coupling can be effectively tuned by the twist angle, the optical injection is expected to be different from both the graphene and bilayer graphene, and it is possible to further understand the effects of interlayer coupling on optical coherent control. This is the focus of this work.

We arrange the paper as follows. In Section~\ref{sec:model} we
introduce an effective model for TBG and expressions for injection
coefficients. In Section~\ref{sec:discuss} we present the main spectra
features  for the injection coefficients at a twist angle $4^\circ$ as
an example and discuss the contributions 
from different electronic states using the electron energy resolved
injection coefficients, then we study the effects of chemical
potentials and the twist angles. We conclude in
Section~\ref{sec:conclusion}.

\section{MODELS\label{sec:model}}
\subsection{Electronic Model Hamiltonian}
TBG with a twist angle $\theta$ can be
formed from rotating the upper and lower layers of AB-stacked bilayer graphene by angles of $-\frac{\theta}{2}$ and $\frac{\theta}{2}$\cite{bistritzer2011moire}, respectively. The primitive reciprocal lattice vectors of the
unrotated graphene layer are chosen as 
\begin{equation}      
  \bm{b_1}=\frac{2\pi}{a_0}\begin{pmatrix}
    \frac{1}{\sqrt{3}}\\ 
    -1
  \end{pmatrix}\,,\quad
  \bm{b_2}=\frac{2\pi}{a_0}\begin{pmatrix}
    \frac{1}{\sqrt{3}}\\ 
    1
  \end{pmatrix}\,,
\end{equation}
then the primitive reciprocal lattice vectors of TBG can be taken as
\begin{align}
  \bm{t_1}&=R\left(-\frac{\theta}{2}\right)\bm{b_2}-R\left({\frac{\theta}{2}}\right)\bm{b_2}\,,\\
  \bm{t_2}&=R\left({\frac{\theta}{2}}\right)\bm{b_1}-R\left({-\frac{\theta}{2}}\right)\bm{b_1}\,,
\end{align}
with the rotation matrix $R(\theta)=$
$\begin{pmatrix}\cos\theta&-\sin\theta\\ \sin\theta&\cos\theta
\end{pmatrix}
$. After the rotation, the Dirac point
$\bm{K}=(\bm{b_1}-\bm{b_2})/3$ of the unrotated layer is
folded to TBG reciprocal space as the Dirac points
$\bm{K_t}=(\bm{t_1}-2\bm{t_2})/3$ from the upper layer and
$\bm{K_t'}=(2\bm{t_1}-\bm{t_2})/3$ from the lower layer; similarly, the Dirac point $\bm{K'}=-\bm{K}$ of the unrotated layer is
folded to TBG reciprocal space as $\bm{K_t'}$ and  $\bm{K_t}$,
respectively. These two valleys are decoupled. The low energy electronic excitations around the $\nu$th
valley ($\nu=+$ for the $\bm K$ point and $\nu=-$ for the $\bm
K^\prime$ point) of each graphene layer can be determined by a
$4\times4$ continuum effective Hamiltonian \cite{bistritzer2011moire}
\begin{equation}
  H^{(\nu)}(\bm{\nabla},\bm{r})=\begin{pmatrix}
    h^{(\nu)}(-\frac{\theta}{2},-i\bm{\nabla}-\nu \bm{K_t})&T^{(\nu)}(\bm{r}) \\ \left[T^{(\nu)}(\bm{r})\right]^\dagger&h^{(\nu)}(\frac{\theta}{2},-i\bm{\nabla}-\nu\bm{K_t'})
  \end{pmatrix}\,,
\end{equation}
where $h^{(\nu)}$ gives the graphene Hamiltonian in the $\nu$th valley as 
\begin{equation}
  h^{(\nu)}(\theta,\bm{k})=\hbar v_f\begin{pmatrix}
    0&e^{i\nu\theta}\left(ik_x+\nu k_y\right) \\
    e^{i\nu\theta}\left(-ik_x+\nu k_y\right)&0
  \end{pmatrix}\,,
\end{equation}
and $T^{(\nu)}(\bm{r})$ describes the interlayer coupling as \begin{equation}
	T^{(\nu)}(\bm{r})=w_0\left(T_1^{(\nu)}+T_2^{(\nu)} e^{-i\nu\bm{t_1}\cdot\bm{r}}+T_3^{(\nu)} e^{-i\nu\bm{t_2}\cdot\bm{r}}\right),
\end{equation}
with 
\begin{equation}
	T_1^{(\nu)}=\begin{pmatrix}1&1\\1&1
        \end{pmatrix}\,,\,
        T_2^{(\nu)}=\begin{pmatrix}e^{i\nu\frac{2\pi}{3}}&1\\e^{-i\nu\frac{2\pi}{3}}&e^{i\nu\frac{2\pi}{3}}
        \end{pmatrix}
      \,,\,
      T_3^{(\nu)}=\begin{pmatrix}e^{-i\nu\frac{2\pi}{3}}&1\\e^{i\nu\frac{2\pi}{3}}&e^{-i\nu\frac{2\pi}{3}}
      \end{pmatrix}\,.
    \end{equation}
The parameter $v_f=\sqrt{3}\gamma_0 a_0/(2\hbar)$ is the Fermi velocity of graphene with $\gamma_0=3$~eV, and $w_0=110~\text{meV}$ is the interlayer coupling
strength. Obviously, the interlayer coupling potential
$T^{(\nu)}(\bm{r})$ is periodic in space with
primitive lattice vectors determined by the primitive reciprocal vectors $\bm{t_1}$ and
$\bm{t_2}$. The continuum model adopted here is appropriate for twist angles less than or equal to $10^\circ$.

The Schr\"odinger equation in the $\nu$th valley becomes 
\begin{equation}
  H^{(\nu)}(\bm{\nabla},\bm{r})\psi^{(\nu)}(\bm{r})=E\psi^{(\nu)}(\bm{r}).
\end{equation}
For a periodic potential, the eigen wavefunctions are Bloch states and they can be expanded in plane waves as 
\begin{equation}
  \psi^{(\nu)}(\bm{r})=\frac{1}{2\pi}e^{i\bm{k}\cdot \bm{r}}\sum_{nm}e^{i\nu \bm t_{nm}\cdot\bm{r}}C_{nm\bm{k}}^{(\nu)},
\end{equation}
with $\bm t_{nm}=n\bm t_1+m\bm t_2$. 
The expansion coefficient $C_{nm\bm{k}}^{(\nu)}$ is a four-component column vector, which can be further written into a compact column vector $C_{\bm{k}}^{(\nu)}$ with elements $[C_{\bm k}^{(\nu)}]_{nm}=C_{nm\bm k}^{(\nu)}$, then the  eigen equation can be written as 
\begin{equation}
  H_{\bm{k}}^{(\nu)} C_{s\bm{k}}^{(\nu)}=\epsilon_{s\bm{k}}^{(\nu)} C_{s\bm{k}}^{(\nu)},
\end{equation}
where the subscript $s$ labels the band, the matrix elements of
$H_{\bm{k}}^{(\nu)}$ between $(n_1m_1)$ and $(n_2m_2)$ is a $4\times 4$ matrix
\begin{align}
    \left[H_{\bm{k}}^{(\nu)}\right]_{n_1m_1,n_2m_2}
  =&\begin{pmatrix} h^{(\nu)}(-\frac{\theta}{2},
            \bm{k}+\nu\bm{t}_{n_1m_1}-\nu\bm{K_t})
            &0\\
            0&h^{(\nu)}(\frac{\theta}{2},\bm{k}+\nu\bm{t}_{n_1m_1}-\nu\bm{K_t'})
          \end{pmatrix}
\delta_{n_1,n_2}\delta_{m_1,m_2}\notag\\
   &+w_0\begin{pmatrix}
            &T_1^{(\nu)}\\
            \left(T_1^{(\nu)}\right)^\ast 
          \end{pmatrix}           
  \delta_{n_1,n_2}\delta_{m_1,m_2}
  +w_0\begin{pmatrix}
    0&T_2^{(\nu)}\delta_{n_1,n_2-1}\\\left(T_2^{(\nu)}\right)^\ast\delta_{n_1,n_2+1}&0
  \end{pmatrix}
\delta_{m_1,m_2}\notag\\
   &+w_0\begin{pmatrix}0&T_3^{(\nu)}\delta_{m_1,m_2-1}\\\left(T_3^{(\nu)}\right)^\ast\delta_{m_1,m_2+1}&0
   \end{pmatrix}
\delta_{n_1,n_2}\,.
\end{align}
The inplane velocity operator is calculated  from $\bm v_{\bm
  k}^{(\nu)}=\hbar^{-1}\bm\nabla_{\bm k}H^{(\nu)}_{\bm k}$, and its 
 matrix elements between band eigenstates are
\begin{equation}
 \bm v_{s_1s_2\bm{k}}^{(\nu)}= \left[C_{s_1
      \bm{k}}^{(\nu)}\right]^\dag \bm v_{\bm
    k}^{(\nu)}C_{s_2\bm k}^{(\nu)}\,.
\end{equation}
The time reversal symmetry connects these two valleys $\nu=\pm$, and it can also be directly verified that
\begin{equation}
	H_{\bm{k}}^{(-)}=\left[H_{-\bm{k}}^{(+)}\right]^\ast.
\end{equation}
Therefore we can always choose
$C_{s\bm{k}}^{(-)}=\left[C_{s(-\bm{k})}^{(+)}\right]^\ast$ and
$\epsilon_{s\bm{k}}^{(-)}=\epsilon_{s(-\bm{k})}^{(+)}$, then the velocity
matrix elements satisfy
\begin{equation}
  \bm v_{s_1s_2\bm{k}}^{(-)}=-\bm v_{s_2s_1(-\bm{k})}^{(+)}.
\end{equation}
\subsection{Carrier Injection and Coherent Current Injection } 
We consider the two-color coherent control of injected carriers and currents in TBG induced by an electric field $\bm{E}\left(t\right)=\bm{E}_\omega e^{-i\omega
  t}+\bm{E}_{2\omega}e^{-2i\omega t}+c.c.$, where the $2\omega$ beam
is usually generated from the second harmonic of the $\omega$
beam. The injection of a physical quantity $P$ can be described\cite{rioux2014interference} by
\begin{align}
  \frac{dP(t)}{dt}
  =&p_1^{ab}(\omega)[E^{a}_\omega]^\ast
     E^b_\omega+
     p_1^{ab}(2\omega)[E^{a}_{2\omega}]^\ast
     E^b_{2\omega}+p_2^{abcd}(\omega)\left[E^{a}_{\omega}\right]^\ast
     [E^b_\omega]^\ast E^c_\omega E^d_\omega\notag\\
		&+\left\{p_{12}^{abc}(\omega)
           [E^{a}_{2\omega}]^\ast E^b_{\omega} E^{c}_{\omega}+c.c\right\}\,.\label{eq:inj}
\end{align}
The first two terms involving $p_1^{ab}\left(\omega\right)$ describe the injection induced by
one-photon absorption at photon frequencies $\omega$ and 2$\omega$,
respectively. The third term involving $p_2^{abcd}\left(\omega\right)$
describes the injection induced by degenerate two-photon
absorption at photon frequency $\omega$. When one-photon
absorption at $2\omega$ and two-photon absorption at $\omega$ occurs simultaneously,
the same electronic states can be optically excited to the same final
electronic states by two different quantum paths, which lead to an
interference, giving the coherent control of the injection. 
All these response coefficients can be derived  from Fermi-Golden rule\cite{rioux2014interference,van2001coherence} and can be written as the sum of the
contributions from two valleys $p=\sum_{\nu=\pm} p^{(\nu)}$ with 
\begin{align}
  p_1^{(\nu);ab}(\omega)
  =&2\times 
     2\pi\left(\frac{e}{\hbar\omega}\right)^2\sum_{ss^\prime}\int\frac{d\bm
     k}{\left(2\pi\right)^2}P_{ss'\bm{k}}^{(\nu)}\left(v_{ss^\prime
     \bm{k}}^{(\nu);a}\right)^\ast v_{ss^\prime
     k}^{(\nu);b}f_{s^\prime sk}^{(\nu)}\delta(\omega_{ss^\prime
     \bm{k}}^{(\nu)}-\omega)\,,\label{eq:p1}\\ 
  p_2^{(\nu);abcd}(\omega)=
   &2\times 
     2\pi\left(\frac{e}{\hbar\omega}\right)^4\sum_{ss^\prime
     }\int\frac{d\bm k}{\left(2\pi\right)^2}P_{ss^\prime
     \bm{k}}^{(\nu)}\left(w_{ss^\prime
     \bm{k}}^{(\nu);ab}\right)^\ast
     w_{ss^\prime\bm{k}}^{(\nu);cd}f_{s^\prime s\bm{k}}^{(\nu)}\delta(\omega_{ss^\prime\bm{k}}^{(\nu)}-2\omega)\,,\label{eq:p2}\\
  p_{12}^{(\nu);abc}(\omega)=
   &-2\times \pi
     i\left(\frac{e}{\hbar\omega}\right)^3\sum_{ss^\prime}\int\frac{d\bm
     k}{\left(2\pi\right)^2}P_{ss^\prime\bm{k}}^{(\nu)}\left(v_{ss^\prime\bm{k}}^{(\nu);a}\right)^\ast
     w_{ss^\prime\bm{k}}^{(\nu);bc}f_{s^\prime s\bm{k}}^{(\nu)}\delta(\omega_{ss^\prime\bm{k}}^{(\nu)}-2\omega)\,.\label{eq:p12}
\end{align}
Here the superscripts $a,b,c,d$ stand for the Cartesian directions
$x,y$. The prefactor $2$ comes from the spin degeneracy.
$f_{s^\prime
  s\bm{k}}^{(\nu)}=f_{s^\prime\bm{k}}^{(\nu)}-f_{s\bm{k}}^{(\nu)}$
gives the electron population difference where
$f_{s\bm{k}}^{(\nu)}=\Theta(\mu-\varepsilon^{(\nu)}_{s\bm{k}})$ is  the Fermi-Dirac distribution at
chemical potential $\mu$ and zero temperature with $\Theta$ being
the Heaviside step function, and the matrix elements $w_{ss^\prime \bm{k}}^{(\nu);ab}$ are given as
\begin{align}
    w_{ss^\prime\bm{k}}^{(\nu);ab}&=\sum_m\frac{v_{sm\bm{k}}^{(\nu);a}v_{ms^\prime\bm{k}}^{(\nu);b}+v_{sm\bm{k}}^{(\nu);b}v_{ms^\prime\bm{k}}^{(\nu);a}}{\omega_{ms\bm{k}}^{(\nu)}+\omega_{ms^\prime\bm{k}}^{(\nu)}-i\gamma}\,,\label{eq:w}
\end{align}
with
$\hbar\omega_{ss^\prime\bm{k}}^{(\nu)} =
\epsilon_{s\bm{k}}^{(\nu)}-\epsilon_{s^\prime\bm{k}}^{(\nu)}$. The quantity $\gamma$ is a phenomenological damping prameter to avoid divergence.

In conventional semiconductors, the absorption processes for
$p_1^{(\nu);ab}(\omega)$, $p_2^{(\nu);abcd}(\omega)$, and
$p_{12}^{(\nu);abc}(\omega)$ can occur only when the photon
energy $2\hbar\omega$ is larger than the band gap $E_g$. {Because TBG has no band gap in
our adopted model}, these processes in principle can
occur at any photon energy in an undoped TBG. However, when TBG is
doped to a chemical potential $\mu$, the population difference
$f_{s^\prime s\bm{k}}^{(\nu)}$ leads to an absorption edge by Pauli
blocking. This is equivalent to a chemical potential induced effective gap
parameter $E_g$, which can set the injection edge. For small $\mu$, it
is the same as graphene {with} $E_g=2|\mu|$.

For carrier injection, we set $P_{ss^\prime \bm k}^{(\nu)}=1$ and use
symbols $\xi_1$ and $\xi_2$ {as} one- and
two-photon injection coefficients; for current injection, we
set $P_{ss^\prime \bm k}^{(\nu)}=e(v_{ss\bm k}^{(\nu);g}-v_{s^\prime
  s^\prime \bm k}^{(\nu);g})$ and use symbols $\eta^{g}$ {as}
injection coefficients. The injection coefficient $\xi_1$ is a second order tensor, $\xi_2$
and $\eta$ are fourth order tensors. 
The crystal symmetry of TBG depends on the initial stacking order and
rotation center\cite{bistritzer2011moire,koshino2018maximally}. In our model, the point group of the structure is $D_6$. Thus the nonzero inplane components\cite{boyd2020nonlinear} are
$\xi_1^{xx}=\xi_2^{yy}$  for $\xi_1^{ab}$, 
$\xi_2^{xxxx}=\xi_2^{yyyy}=\xi_2^{xxyy}+2\xi_2^{xyxy},\xi_2^{xyxy}=\xi_2^{xyyx}$
for $\xi_2^{abcd}$, and similar results for $\eta_{12}^{gabc}$. By
inspecting the expression in Eqs.~(\ref{eq:p1})-(\ref{eq:p12}), it can
be found that $\xi_1^{ab}=[\xi_1^{ba}]^\ast$ and
$\xi_2^{abcd}=[\xi_2^{cdab}]^\ast$, thus $\xi_1^{xx}$, $\xi_2^{xxyy}$,
and $\xi_2^{xyxy}$ are all real, and $\eta_{12}^{xxyy}$ and
$\eta_{12}^{xyxy}$ are in general complex. Furthermore, there
exists time reversal symmetry linking the $\pm$ valleys, then all
injection coefficients can be obtained from the calculation of one
valley. For
one-photon carrier injection  it gives
$\xi_1^{(+);ab}(\omega)=\xi_1^{(-);ab}(\omega)$; for two-photon
carrier injection and coherent current injection, we have
\begin{align}
  \xi_2^{abcd}(\omega)=
  &8\pi\left(\frac{e}{\hbar\omega}\right)^2\sum_{ss^\prime
   }\int\frac{d\bm k}{\left(2\pi\right)^2}f_{s^\prime s\bm{k}}^{(+)}\delta(\omega_{ss^\prime\bm{k}}^{(+)}-2\omega)\notag\\
 &\times \text{Re}\left[\left(w_{1;ss^\prime
       \bm{k}}^{(+);ab}\right)^\ast
   w_{1;ss^\prime\bm{k}}^{(+);cd}+\left(w_{2;ss^\prime
       \bm{k}}^{(+);ab}\right)^\ast
   w_{2;ss^\prime\bm{k}}^{(+);cd}\right]\,,\label{eq:xi21} \\
  \eta_{12}^{gabc}(\omega)
  =&-4\pi\left(\frac{e}{\hbar\omega}\right)^3
     \sum_{ss^\prime}\int\frac{d\bm k}{(2\pi)^2}f_{s^\prime sk}^{(+)} \delta(\omega_{ss^\prime\bm{k}}^{(+)}-2\omega) \notag\\
  &\times \left\{-i\text{Re}[(v_{ss\bm{k}}^{(+);g}-v_{s^\prime
     s^\prime\bm{k}}^{(+);g})v_{s^\prime
     s\bm{k}}^{(+);a}w_{1;ss^\prime\bm
     k}^{(+);bc}]+\text{Im}[(v_{ss\bm{k}}^{(+);g}-v_{s^\prime s^\prime
     \bm{k}}^{(+);g})v_{s^\prime
     s\bm{k}}^{(+);a}w_{2;ss^\prime}^{(+);bc}]\right\}\,,\label{eq:eta12}
\end{align}
with $w_{ss^\prime\bm k}^{(\nu);ab}=w_{1;ss^\prime\bm
  k}^{(\nu);ab}+i w_{2;ss^\prime\bm k}^{(\nu);ab}$ and 
\begin{align} 
  w_{1;ss^\prime\bm k}^{(\nu);ab}
  &=\sum_m\frac{v_{smk}^{(\nu);a}v_{ms'\bm{k}}^{(\nu);b}+v_{sm\bm{k}}^{(\nu);b}v_{ms'\bm{k}}^{(\nu);a}}{\left(\omega_{ms\bm{k}}^{(\nu)}+\omega_{ms'\bm{k}}^{(\nu)}\right)^2+\gamma^2}(\omega_{ms\bm{k}}^{(\nu)}+\omega_{ms'\bm{k}}^{(\nu)})\,,\\
    w_{2;ss^\prime\bm k}^{(\nu);ab}
  &=\sum_m
    \frac{v_{sm\bm{k}}^{(v);a}v_{ms'\bm{k}}^{(v);b}+v_{sm\bm{k}}^{(v);b}v_{ms'\bm{k}}^{(v);a}}{\left(\omega_{ms\bm{k}}^{(\nu)}+\omega_{ms'\bm{k}}^{(\nu)}\right)^2+\gamma^2}\gamma\,.\label{eq:w2}
\end{align}

Here we give a brief discussion on the parameter $\gamma$. For graphene which has only two bands in the simplest model,
such parameter can be taken as $0$ directly because the denominator usually
does not go to zero. However, due to the existence of many bands in
TBG, the denominator in Eq.~(\ref{eq:w}) can go to zero for zero $\gamma$ when
there exists an intermediate state in the middle of the initial and final states,
{\it i.e.},
$\epsilon_{s\bm{k}}^{(\nu)}+\epsilon_{s^\prime\bm{k}}^{(\nu)}
=2\epsilon_{m\bm{k}}^{(\nu)}$. This condition leads to the resonant
one-photon transition at $\omega$ between the initial and intermediate
states; thus it requires the photon energy $2\hbar\omega>E_g$. To
better understand how such divergences affect the injection
coefficients, we show the limit of $\gamma\rightarrow0$ in Eq.~(\ref{eq:w2}) as
\begin{equation}
  \lim\limits_{\gamma\to0}w_{2;ss^\prime\bm{k}}^{(\nu);ab}=\pi\sum_m(v_{sm\bm{k}}^{(\nu);a}v_{ms^\prime\bm{k}}^{(\nu);b}+v_{sm\bm{k}}^{(\nu);b}v_{ms^\prime\bm{k}}^{(\nu);a})\delta(\omega_{ms\bm{k}}^{(\nu)}+\omega_{ms^\prime\bm{k}}^{(\nu)})\,.\label{eq:limit}
\end{equation}
It gives an additional $\delta$ function. In the calculation of
$\eta_{12}^{gabc}(\omega)$, the product of two $\delta$ functions
$\delta(\omega_{ss^\prime\bm{k}}^{(\nu)}-2\omega)\delta(\omega_{ms\bm{k}}^{(\nu)}+\omega_{ms'\bm{k}}^{(\nu)})$
can behave well after integrating over two dimensional wave vector
$\bm{k}$; while, in the calculation of $\xi_2^{abcd}(\omega)$, an
additional product
$\delta(\omega_{ss'\bm{k}}^{(\nu)}-2\omega)[\delta(\omega_{ms\bm{k}}^{(\nu)}+\omega_{ms'\bm{k}}^{(\nu)})]^2$
appears and becomes divergent. However, this is physically meaningful: with the existence of the
one-photon absorption, two-photon absorption is a signature towards
saturated absorption. In this case a finite $\hbar\gamma=10~\text{meV}$ is
adopted unless other value is specified, but one keeps in mind that the results strongly depend on the value of $\gamma$.

\section{Results and discussion\label{sec:discuss}}
In this work, we focus on the injection coefficients for twist
angles between $3^\circ$ and $10^\circ$. In the diagonalization of the Hamiltonian, the plane wave used for the wave function expansion are taken as $e^{i\nu \bm t_{nm}\cdot\bm r}$ with $nm\in[-N,N][-N,N]$; and the numerical calculation of the injection coefficients is performed by discretizing the TBG Brillouin zone in a $M\times M$ grid. The delta function is approximated by a Gaussian function with an energy broadening $20~\text{meV}$. The convergences of the results are checked with the values of $N$ and $M$. At the twist angle $4^\circ$, we choose $N=5$ and $M=150$.

\subsection{Band structure and Density of states}
\begin{figure*}[h]
   \centering
   \includegraphics[width=\linewidth]{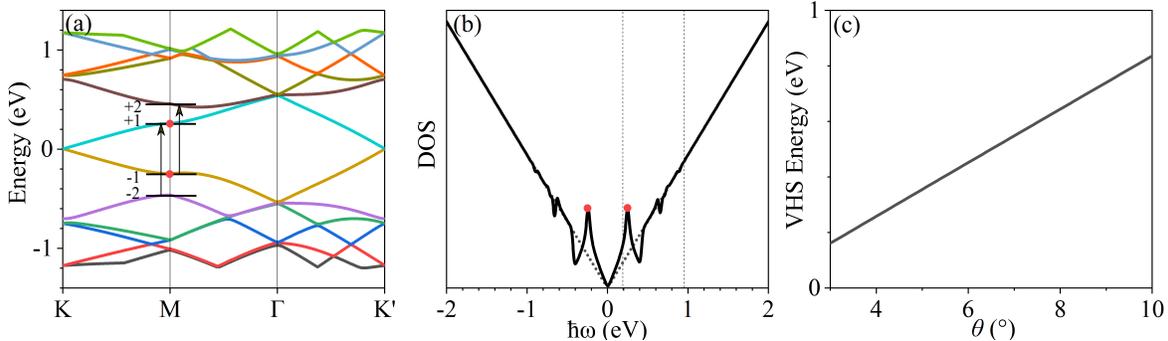}
   \caption{ Illustration of the (a) band structure and (b) DOS for twist angle $\theta=4^\circ$. The two dotted vertical lines
     separate the three regimes in positive energies. (c) The twist angle dependence of energy at VHS.} 
   \label{fig:band} 
 \end{figure*}

 As an example, we plot the band structure at $\theta=4^\circ$ in
 Fig.~\ref{fig:band} (a). The bands are labelled by
 $s=\pm1,\pm2,\cdots$, where negative/positive $s$ are for bands with
 energies below/above zero. Together with the DOS shown
 in Fig.~\ref{fig:band} (b), the band structure clearly shows three energy regimes: (1) There exists a linear
 regime of the $\pm1$ band around the Dirac points, which has already
 been well discussed and characterized by a renormalized Fermi velocity\cite{bistritzer2011moire}
 $v_f^\prime=\frac{1-3\alpha^2}{1+6\alpha^2}v_f$, with  $\alpha=3a_0w_0/(8\pi v_f\sin({\theta}/{2}))$. It is easy to
conclude that the physics in this regime should be similar to that of
 graphene, but with a smaller Fermi velocity and a larger DOS. At
 $\theta=4^\circ$, this regime is about in the energy range
 $[-0.19, 0.19]$~eV. (2) When the electron energies are higher than $0.95~\text{eV}$, the interlayer coupling shows little effect on
 the DOS. This is easy to understand because the coupling energy
 $w_0=110$~meV only contributes a bit to the electron energy in this
 case. The optical response of these
 two regimes is  similar to the results of graphene. Because the continuum model for graphene is appropriate for electronic states in the linear dispersion regime(usually $<\gamma_0$),  our calculation is performed for the photon energy less than $5~\text{eV}$. (3) When the energy is between $0.19~\text{eV}$ and $0.95~\text{eV}$,
 the band structure is complicated compared to that of
 graphene. There are multiple bands in this regime, and the M
 points of the $\pm$1 bands (shown as red points in
 Fig.~\ref{fig:band} (a,b)) are saddle 
 points, leading to VHS in the DOS at
 an energy about $0.25$~eV. The energies of these saddle points are
 approximately linear with the twist angle, as shown in
 Fig.~\ref{fig:band} (c). When the electron energy exceeds the VHS, there is a sudden decrease of the DOS, because the states are shifted to the M
 points to form VHS\cite{moon2013optical}.

\subsection{Injection coefficients at $4^\circ$}
To clearly show how the interlayer coupling at different twist
angles affect the injection, it is constructive to consider the ratio
between the obtained results and those results for uncoupled TBG, where the interlayer coupling strength is set as $w_0=0$. The uncoupled TBG becomes simply two
uncoupled monolayer of graphene, and the injection occurs in each layer
only, for which analytic results have been obtained
\cite{rioux2011current}. The injection in graphene is
isotropic with additional relations for the injection coefficients as $\xi_2^{xyxy}=-\xi_2^{xxyy}$ and
$\eta_{12}^{xyxy}=-\eta_{12}^{xxyy}$. Thus in an uncoupled TBG for any
twist angle $\theta$ the injection coefficients are just twice of those
in graphene, as 
\begin{align}
  \xi_1^{xx}(\omega)|_u=&\frac{e^2}{\hbar^2\omega}\,, \quad 
	\xi_2^{xxxx}(\omega)|_u=\frac{2\hbar v_f^2e^4}{(\hbar\omega)^5}\,,\quad
	\eta_{12}^{xxxx}(\omega)|_u=i\frac{v_f^2e^4}{(\hbar\omega)^3}\,.\label{eq:resgh}
\end{align}
The normalized injection coefficients are defined as
$\overline{\xi}_1^{ab}(\omega)=\xi^{ab}_1(\omega)/\xi_1^{xx}(\omega)|_u$,
$\overline{\xi}_2^{abcd}(\omega)=\xi_2^{abcd}(\omega)/
\xi_2^{xxxx}(\omega)|_u$, and 
$\overline{\eta}_{12}^{gabc}(\omega)=\eta_{12}^{gabc}(\omega)/\eta_{12}^{xxxx}(\omega)|_u$,
which are dimensionless quantities. In an uncoupled TBG, $\eta_{12}^{xxxx}(\omega)|_u$ is a pure imaginary number; with
 the inclusion of interlayer coupling, $\eta_{12}^{gabc}$ is in general
complex, but the calculations show its real part is 2 orders of
  magnitude smaller than its imaginary part, which will be presented below. 

\begin{figure*}[h]
  \centering
  \includegraphics[width=\linewidth]{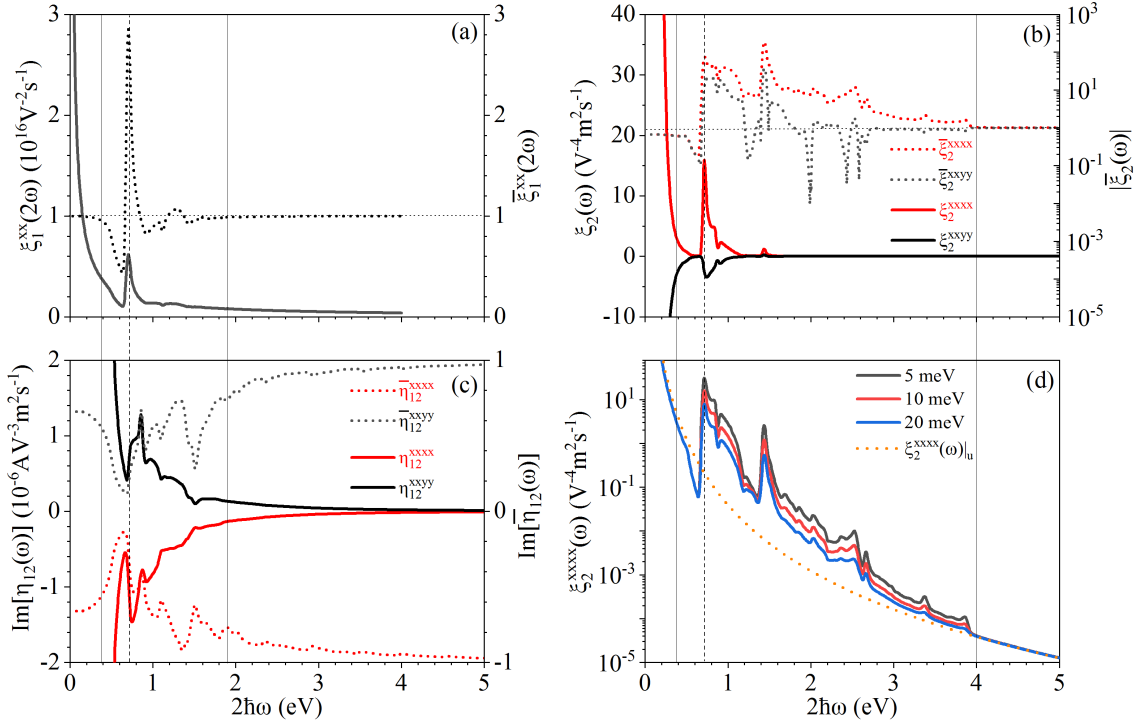}
  \caption{ Spectra of injection  coefficients of TBG with $\theta=4^\circ$ and $\mu=0$. (a) One-photon
    carrier injection coefficient $\xi_1^{xx}(\omega)$, (b) two-photon
    carrier injection coefficients  $\xi_2^{xxxx}(\omega)$ and
    $\xi_2^{xxyy}(\omega)$, (c) the imaginary part of coherent current
    injection coefficients  $\eta_{12}^{xxxx}(\omega)$ and
    $\eta_{12}^{xxyy}(\omega)$. In (a,b,c) the right axis
    gives the normalized injection coefficients. The
    vertical solid lines located at $2\hbar\omega=0.38, 1.9$ or $4~\text{eV}$ divide
    the whole spectra into three regimes. The vertical dashed lines indicate the energy for the transition at the M point from band $-1$ ($-2$) to $+2$ ($+1$).
    (d) $\xi^{xxxx}_2(\omega)$ at
    different $\hbar\gamma=5, 10, 20~\text{meV}$. }
  \label{fig:injectspectra}
\end{figure*}
Figure~\ref{fig:injectspectra} gives the spectra of these coefficients
for a twist angle $4^{\circ}$ at zero chemical
potential. Corresponding to three regimes of the band structure, the spectra of injection coefficients can also be divided into three regimes, which are separated by $2\hbar\omega=0.38~\text{eV}$ and $1.9~\text{eV}$ for $\xi_1^{ab}$ and $\eta_{12}^{gabc}$ or $2\hbar\omega=0.38~\text{eV}$ and $4~\text{eV}$ for $\xi_2^{abcd}$. The second boundary is different due to the existence of unique divergence in two-photon absorption.  At low energy regime
$2\hbar\omega<0.38~\text{eV}$, the injection occurs mostly between the {bands}
$\pm$1 in the linear dispersion regime, the results are similar to those
of graphene but with a smaller Fermi velocity $v_f^\prime$.
From Eq.~(\ref{eq:resgh}), the one-photon
injection coefficient is the same as the uncoupled TBG because it is
independent of the Fermi velocity; however,
two-photon carrier injection and two-color coherent current injection show
smaller coefficients due to the smaller Fermi velocity. These features are
clearly shown by the values  $\overline{\xi}_1^{xx}=1$,
$\overline{\xi}_2^{xxxx}$ and $\overline{\eta}_{12}^{xxxx}$ which are
approximately constant but less than $1$. In the high energy regime with photon energies 
$2\hbar\omega>1.9$~eV for $\xi^{xx}_1(\omega)$ and $\eta^{xxxx}_{12}(\omega)$, or $2\hbar\omega>4~\text{eV}$ for $\xi^{xxxx}_2(\omega)$, the coefficients gradually approach
the case without interlayer coupling.

In the middle regime, the spectra of injection coefficients contain
fruitful fine structures, which are also clearly shown in the normalized injection coefficients.  All of them have dips  at $2\hbar\omega\approx0.68~\text{eV}$, which are induced by the optical transitions involving
the states with smaller DOS shown in Fig.~\ref{fig:band} (b). At $2\hbar\omega\approx0.72~\text{eV}$,  $\bar\xi_1^{xx}(\omega)$ appears a peak with a value around 3, which arises from the optical transitions at the VHS. However, both $\bar\eta^{xxxx}_{12}(\omega)$ and $\bar\eta^{xxyy}_{12}(\omega)$ show local peaks at $2\hbar\omega=0.76~\text{eV}$ and $0.85~\text{eV}$, respectively. The higher photon energies of these peak
locations are because the VHS has zero carriers velocity $v_{ss\bm k}^g-v^g_{s^\prime s\prime\bm k}$, which lowers and shifts the peak for current injection coefficients. After the peak, the injection coefficients decrease with
some fine features (tiny dips), which are induced by the existence of multiple
bands. In the whole middle regime, the normalized current injection coefficients are less than 1.

Different from one-photon carrier injection and two-color
coherent injection, where the injection coefficients are at the same
order of magnitude of graphene, the two-photon carrier injection can
be a few hundreds times larger than that of graphene for
photon energies $0.72<2\hbar\omega<4~\text{eV}$. This is because for these
photon energies, both the resonant one-photon optical transition and
resonant two-photon optical transition can exist simultaneously, which
leads to a double resonance discussed after Eq.~(\ref{eq:limit}). To
better illustrate its dependence on $\gamma$, we also plot
$\xi_2^{abcd}(\omega)$ in Fig.~\ref{fig:injectspectra} (d) for
different $\hbar\gamma=5, 10, 20~\text{meV}$. Our calculation indicates that
 the injection coefficients for the double resonant
transitions are approximately proportional to $\gamma^{-1}$.
\begin{figure}[h]
  \centering  
  \includegraphics[width=\linewidth]{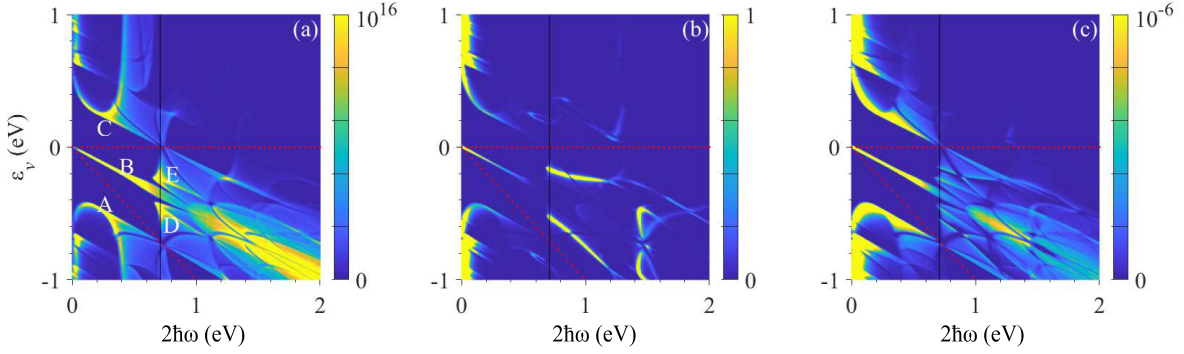}
  \caption{Electron energy resolved injection coefficients (a)
    $\left(\frac{2\hbar\omega}{1\text{eV}}\right)^2\widetilde{\xi}_1^{xx}(\epsilon_v,2\omega)$,
    (b) $\left(\frac{\hbar\omega}{1\text{eV}}\right)^4\widetilde{\xi}_2^{xxxx}(\epsilon_v,\omega)$, and (c)
    $\left(\frac{2\hbar\omega}{1\text{eV}}\right)^3\text{Im}[\widetilde{\eta}_{12}^{xxxx}(\epsilon_v,\omega)]$. The two dotted
    lines give $\epsilon_v=\mu$ and $\epsilon_v=\mu-\hbar\omega$ for
    $\mu=0~\text{eV}$.  }
  \label{fig:res}
\end{figure}
\subsection{Electron energy resolved injection coefficients at $\theta=4^\circ$}
The injection processes can be better understood by using the electron energy resolved injection coefficients, from which the contributions from different electron energies can be visualized.  As an example, for the one-photon injection, it is defined as
\begin{align}
  \widetilde{\xi}_1^{(\nu);ab}(\epsilon_v, 2\omega)
  =&4\pi\left(\frac{e}{\hbar\omega}\right)^2\sum_{ss^\prime}\int\frac{d\bm
     k}{\left(2\pi\right)^2}\left(v_{ss^\prime
     \bm{k}}^{(\nu);a}\right)^\ast v_{ss^\prime
     k}^{(\nu);b}\delta(\varepsilon_{s^\prime\bm
     k}-\epsilon_v)\hbar\delta(\varepsilon_{s\bm
     k}-\epsilon_v-2\hbar\omega)\,,\label{eq:p1res}
\end{align}
from which the injection coefficients can be obtained as
\begin{align}
  \xi_1^{(\nu);ab}(2\omega) = \int d\epsilon_v
  \left[\Theta(\mu-\epsilon_v)-\Theta(\mu-2\hbar\omega-\epsilon_v)\right]
  \widetilde{\xi}_1^{(\nu);ab}(\epsilon_v,  2\omega)\,.\label{eq:xia}
\end{align}
Similar definitions can be applied to get $\widetilde{\xi}_2^{dabc}(\epsilon_v, \omega)$
and $\widetilde{\eta}_{12}^{gabc}(\epsilon_v, \omega)$. Equation~(\ref{eq:xia}) shows that
only the electrons states with energies $\mu-2\hbar\omega<\epsilon_v<\mu$
contribute to the injection process. For uncoupled TBG or for graphene, the injection occurs as
$\epsilon_v=-\hbar\omega<-|\mu|$. 

Figure \ref{fig:res} shows electron energy resolved injection coefficients
$\widetilde{\xi}_{1}^{xx}(\epsilon_v,2\omega)$,
$\widetilde{\xi}_2^{xxxx}(\epsilon_v,\omega)$, and
$\widetilde{\eta}_{12}^{xxxx}(\epsilon_v,\omega)$. Taking
$\widetilde{\xi}_{1}^{xx}(\epsilon_v,2\omega)$ as an example, the main contributions locate in
the regions identified by the optical transitions
between bands A: $-2\to-1$, B: $-1\to1$, C: $1\to2$, D: $-2\to1$,
E: $-1\to2$, and F: for transitions from or to higher energy bands. The regions D and E include the optical transitions around the VHS. At $\mu=0$, the electron energy resolved injection
coefficients mostly locate in the region $-2\hbar\omega<\epsilon_v<0$ given by the dashed red lines, including the regions B, E, and D. For nonzero $\mu$,  contributions from other regions can be
tuned on or off, which will be discussed in next section.

\subsection{Chemical potential dependence}
\begin{figure*}[h]
  \centering
  \includegraphics[width=\linewidth]{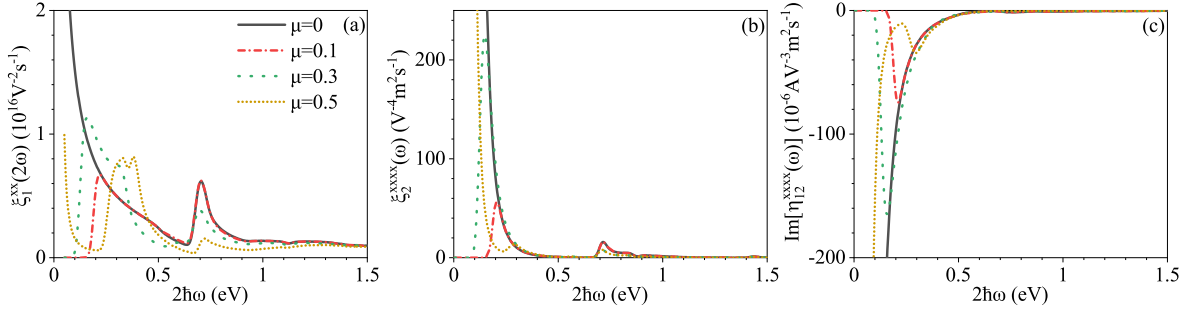}
  \caption{Spectra of injection coefficients (a)
     $\xi_1^{xx}(2\omega)$, (b) $\xi_2^{xxxx}(\omega)$, (c) 
   $\text{Im}[\bar\eta_{12}^{xxxx}(\omega)]$
    for chemical potentials $\mu$=$0, 0.1,
    0.3, 0.5~\text{eV}$.}
  \label{fig:mu}
\end{figure*}
Now we consider how doping affects the injection
coefficients. Figure~\ref{fig:mu} shows the spectra for $\mu=0, 0.1,
0.3, 0.5~\text{eV}$ at $4^\circ$. For $\mu=0.1~\text{eV}$, the
chemical potential is in the linear dispersion regime. Following the
results of graphene, the chemical potential  induced  effective band gap is $E_g=2|\mu|$, thus all injection coefficients show an onset energy at $2\hbar\omega=0.2~\text{eV}$. The new transitions from the
 $+1$ band to higher bands (from region C in Fig. 3) require higher photon
energies and the contribution is negligible, so the results after the effective gap are almost the same as those results at zero chemical potential. When the
chemical potential increases to $0.3~\text{eV}$, which still lies in the +1 band, a direct consideration of the effective gap should be as high as $0.6~\text{eV}$. However,  the higher doping level makes transitions from the +1 band to higher bands (from the region C in Fig. 3) requires less photon energy, which reduces the effective gap  to $0.15~\text{eV}$. When the chemical potential is
$0.5~\text{eV}$, the onset energy goes to even smaller around 0. From
Fig. 3, all injections contributed from the regions B, D, and E are suppressed, but region C contributes greatly for small photon energies,  and enhances the one-photon carrier injection for photon energies between $0.27~\text{eV}$ and $0.31~\text{eV}$, and two-photon carrier injection and coherent current injection for photon energies between $0.27~\text{eV}$ and $0.5~\text{eV}$.

\subsection{Twisted angle dependence}
\begin{figure*}[h]
  \centering
  \includegraphics[width=\linewidth]{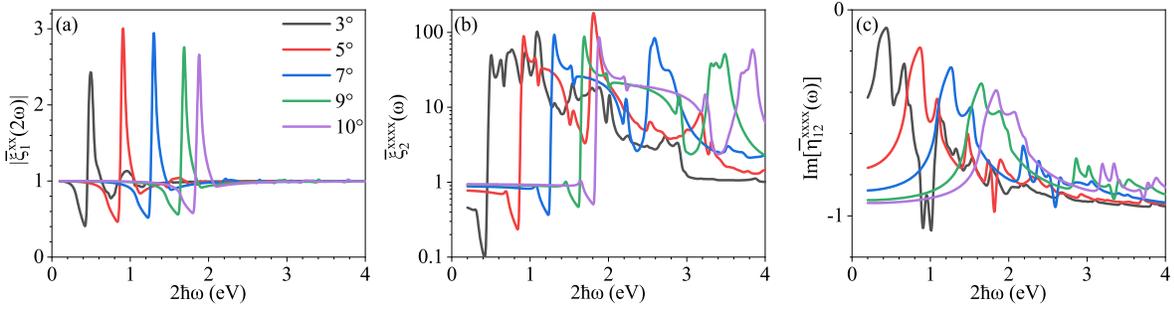}
  \caption{Photon energy dependence of the normalized injection
    coefficients (a) $\overline{\xi}_1^{xx}(\omega)$,
    (b) $\overline{\xi}_2^{xxxx}(\omega)$, and
    (c) $\text{Im}[\eta_{12}^{xxxx}(\omega)]$ for different twist
    angles $3^\circ$, $5^\circ$, $7^\circ$, $9^\circ$, and $10^\circ$.}
  \label{fig:theta}
\end{figure*}
Figure~\ref{fig:theta} gives the normalized injection coefficients for $\theta=3^\circ$, $5^\circ$,
$7^\circ$, $9^\circ$ and $10^\circ$. For $\theta>10^\circ$, the continuum model for electronic states is not suitable \cite{bistritzer2011moire}.  The spectra at different
twist angles show very similar features to that of $4^\circ$.  With
increasing $\theta$, the peaks/valleys are shifted to larger photon
energies, which is consistent with our analysis of the band
structure. These results indicate that optical injection processes can
be effectively tuned by the twisted angle.

\section{CONCLUSION\label{sec:conclusion}}

We have theoretically investigated one- and two-photon carrier
injection and two-color coherent current injection in twisted bilayer
graphene for twist angles between $3^\circ$ and 
$10^\circ$, where the injection coefficients are numerically evaluated
at zero temperature for different chemical potentials. Compared to the
results for graphene, the spectra of injection coefficients in twisted
bilayer graphene exhibit different features in three energy regimes:
in the low energy regime where the band structure is approximately
linear, all injection coefficients have similar behaviors as that of
graphene, but with a different amplitude determined by the
renormalized Fermi velocity; for very high  photon energies, all
injection coefficients are almost the same as that of graphene,
because the involved electronic states have large energies which are
not  effectively affected by the 
interlayer coupling; and in the middle regime, the carrier
injection coefficients show resonant peaks around the Van Hove
Singularity, while the  
current injection coefficients are smaller because the injected
carriers around the Van Hove singularity have 
zero velocity. All these results are characterized from an electron
energy resolved injection coefficients. Due to the existence of
multiple bands, the degenerate two-photon optical transition processes
can have double resonant optical transitions between the initial, the
intermediate, and the final states, which lead to a divergent
two-photon carrier injection and a finite two-color current
injection. 

With the decrease of the twist angle, these features shift to low
photon energies,  suggesting twist angle tunable applications in
far infrared or Terahertz wavelength. The optical injection for very
small  twist angles are not investigated in this work, mostly due to
the difficulties in the numerical calculation that require a large
number of plane wave expansion to get  accurate band
eigenstates. However, at these angles, the strong carrier-carrier
interaction may lead to a different behavior of the optical
injection\cite{bhat2005excitonic}, which is worth for a future
exploration.  

This work mostly focuses on the comparison of the injection coefficients of twisted bilayer graphene and those of graphene, and there also exists other important injection processes appearing only in twisted bilayer graphene that are worth to be explored but not discussed here. For example, the stacking of two layers of graphene opens the responses along the perpendicular direction\cite{gao2020tunable}, which are usually ignored in a monolayer graphene, the two-color optical injection tensors have nonzero components for oblique incident light. Moreover, due to the lower symmetry, a single color light with an appropriate polarization is also possible to inject currents. 

\acknowledgements
This work has been supported by Scientific research project of the Chinese Academy of Sciences Grant No. QYZDB-SSW-SYS038, National Natural Science Foundation of China Grant No. 11774340, 12034003, 12004379 and 11804334. J.L.C. acknowledges the support from Talent Program of CIOMP.

\bibliographystyle{apsrev4-1}
%

\end{document}